\documentclass[conference]{IEEEtran}
\IEEEoverridecommandlockouts 
\usepackage{graphicx}
\usepackage{latexsym}
\usepackage{amsmath}
\usepackage{amsthm}
\usepackage{amssymb,amsmath}
\usepackage{epstopdf}
\usepackage{subfigure}
\date{}
\usepackage{fancyhdr}
\usepackage[font=scriptsize]{caption}

\usepackage{eso-pic}
\definecolor{red}{rgb}{0.8,0.1,0.1}

\begin{document}
\title{Optimal Local Thresholds for Distributed Detection in Energy Harvesting Wireless Sensor Networks}
\author{\IEEEauthorblockN{Ghazaleh Ardeshiri,
		Hassan Yazdani,
		Azadeh Vosoughi~\IEEEmembership{Senior Member,~IEEE}}
	\IEEEauthorblockA{University of Central Florida\\  Email:gh.ardeshiri@knights.ucf.edu, h.yazdani@knights.ucf.edu, azadeh@ucf.edu} }
\maketitle
\begin{abstract}
We consider a wireless sensor network, consisting of $K$ 
heterogeneous 
sensors and a fusion center (FC), that is tasked with solving a binary distributed detection problem. 
Each sensor is capable of harvesting and storing energy 
%energy and storing it in a finite-capacity battery, 
for communication with the FC. 
%The orthogonal communication channels between the sensors and the FC is subject to fading and noise. 
%
For energy efficiency, %during an observation period,
a sensor transmits only if the sensor test statistic exceeds a local threshold $\theta_k$, its channel gain exceeds a minimum threshold, and its battery state can afford the transmission.
Our proposed transmission model at each sensor is motivated by the channel inversion power control strategy in the wireless communication community.
Considering a constraint on the average energy of transmit symbols, 
we study the optimal $\theta_k$'s that optimize two detection performance metrics: %For this optimization, 
%We consider two metrics: 
(i) the detection probability $P_D$ at the FC, assuming that the FC utilizes the optimal fusion rule based on Neyman-Pearson optimality criterion, and (ii) Kullback-Leibler distance (KL) between the two distributions of the received signals at the FC conditioned by each hypothesis.
Our numerical results indicate that $\theta_k$'s obtained from maximizing the KL distance are near-optimal. Finding these thresholds is 
computationally efficient, as it requires only $K$ one-dimensional searches, as opposed 
to a $K$-dimensional search required to find the thresholds that maximize $P_D$.
%, in the sense that they provide similar detection probability to those thresholds obtained . 
\end{abstract}
\IEEEpeerreviewmaketitle
%%%%%%%%%%%%%%%%%%%%%%%%%%%%%%%%%%%%%%%%%%%%%%%%%%%%%%%%%%%%%%%%%%%%%%%%%%%%%%%%%%%%%%%%%%%%%%%%%%%%%%%%%%%%%%%%%%%%%%%%%%%%%%%%%
\vspace{-0.2cm}
\section{Introduction}
\vspace{-0.2cm}
%A wireless sensor network, consisting of a network of sensors with embedded capabilities of sensing, computation, and communication, is used to sense and collect data for a wide range of applications \cite{1}. 
The designs of wireless sensor networks to perform the task of distributed detection are often based on the conventional battery-powered sensors, leading into designs with a short lifetime, due to battery depletion \cite{1,2,3}. 
%A 
Recently, energy harvesting, which can collect energy from renewable resources in ambient environment (e.g., solar, wind, and geothermal energy)
%, energy from radio-frequency signal) 
%or from man-made sources via wireless energy transfer, 
has attracted much attention \cite{4,5}. 
%In addition, energy may be harvested from man-made sources via wireless energy transfer, where energy is transferred from one node to another in a controlled manner \cite{4}.  
Energy harvesting technology in wireless sensor networks promises a self-sustainable system with a lifetime that is not limited by the lifetime of the conventional batteries \cite{2,6,7}. 
%However, it poses new challenges related to the management of the harvested energy, since that the amount of energy available at a sensor is random \cite{2,6}.

In this paper, we consider the distributed detection of a known signal using a wireless network with $K$ energy harvesting sensors and a fusion center (FC). Each sensor makes a noisy observation, corrupted by both additive and multiplicative observation noises.
Each sensor applies an energy detector, to compare its test statistic against a local decision threshold $\theta_k$ (to be optimized), and transmits only if the test statistic exceeds $\theta_k$, its channel gain exceeds a minimum threshold $\zeta_k$, and its battery state can afford transmission. Given our transmission and battery state models, our goal is to investigate the optimal $\theta_k$'s that optimize the detection performance metric, subject to average transmit symbol energy constraint.
The paper organization follows: in Section \ref{section-2} we present our system model, including our transmission and battery state models.
In Section \ref{Section-3} we derive the optimal fusion rule and its corresponding detection and false alarm probabilities $P_D,P_F$, we provide two approximate expressions for the total Kullback-Leibler (KL) distance $KL_{tot}$ at the FC, and we discuss finding $\zeta_k$'s based on the average transmit symbol energy constraint. Section \ref{last-section} illustrates our numerical results on optimizing $\theta_k$'s based on maximizing $P_D$ and $KL_{tot}$, and our concluding remarks. 
%%%%%%%%%%%%%%%%%%%%%%%%%%%%%%%%%%%%%%%%%%%%%%%%%%%%%%%%%%%%%%%%%%%%%%%%%%%%%%%%%%%%%%%%%%%
\section{Our system model and problem Statement}\label{section-2}
\begin{figure}[t]
		\vspace{-2mm}
		\centering
        \includegraphics[width=80mm]{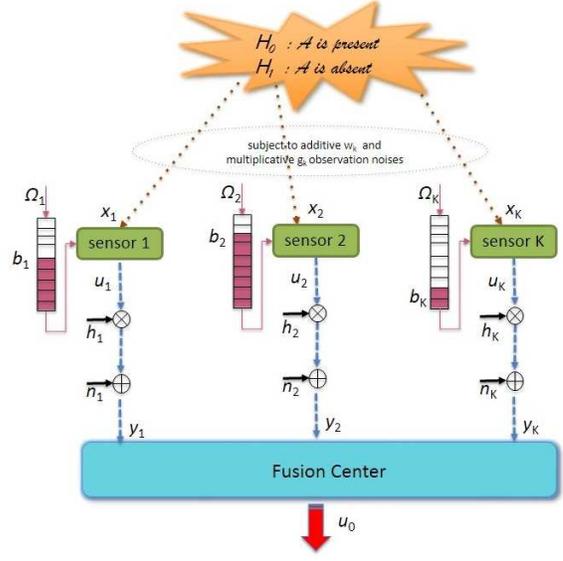}
		\caption{Our System model}
\end{figure}
%
%%
%\begin{figure}
%\centering
    %\subfigure[]
    %{
        %\includegraphics[width=38mm]{Doc1_1.pdf}\label{1}
        %\centering
    %}
    %\subfigure[]
    %{
        %\includegraphics[width=25mm]{Doc2.pdf}\label{2}
        %\centering
    %}
    %\qquad
    %\caption{Our System model during one observation period.}
    %\label{fig:foobar}
    %\vspace{-.6cm}
%\end{figure}
%------------------------------------------------------------------------------------------
We consider a distributed binary hypothesis testing problem where $K$ sensors and a FC are tasked with solving a binary hypothesis testing problem. The particular detection problem we focus on is determining the presence or absence of a known scalar signal ${\cal A}$ (see Fig.\ref{1}). 
%Sensors are in homogeneous in the sense that they have different signal statistics. 
Let $x_k$ denote the local observation at sensor $k$ during an observation period. We assume the following signal model
\vspace{-1mm}
%------------------------------------------------------------------------------------------
%\begin{equation}\label{xk}
%x_{k} =
%\begin{cases}
%g_{k}{\cal A}+w_{k} & ~~~~~~\mathcal{H}_{1} \\
%w_{k} &~~~~~~\mathcal{H}_{0}
%\end{cases} 
%\end{equation}
%
\begin{equation}\label{xk}
\mathcal{H}_{1}: x_{k} ={\cal A} g_{k}+w_{k}, ~~~\mathcal{H}_{0}: x_k=w_k
\vspace{-1mm}
\end{equation}
%
%------------------------------------------------------------------------------------------
%where $s_{k}$ represents the $k^{th}$ signal and $w_{k}$  is the additive Gaussian noise with mean 0 and $\sigma_{w_{k}}^{2}$. The observation noise is assumed to be independent over time and among sensors. 
%\\$g_{k}$ is the fading channel gain of source to sensor $k$ links, and it is statistical CSI, where $|g_k|^2$ has exponential distribution with parameter $\gamma_{g_{k}}$.
%Here, we assume that the absence or the presence of the signal under $\mathcal{H}_{0}$ or $\mathcal{H}_{1}$ is described as
where $w_k$ and $g_k$ are additive and multiplicative observation noises, respectively. We assume $w_k \! \sim \! {\cal N}(0,\sigma_{w_k}^2)$, $g_k \! \sim \! {\cal N}(0,\gamma_{g_{k}})$ and
all observation noises are independent over time and among $K$ sensors. 
%------------------------------------------------------------------------------------------
%\begin{equation}\label{eq2}
%s_{k} =
%\begin{cases}
%A &~~~~~~~~~~: \mathcal{H}_{1} \\
%0 &~~~~~~~~~~: \mathcal{H}_{0}
%\end{cases} 
%\end{equation}
%where $A$ is known scaler.
%------------------------------------------------------------------------------------------
%------------------------------------------------------------------------------------------
During each observation period, sensor $k$ takes $N$ samples of $x_k$ to measure the received signal energy and applies an energy detector to make a binary decision, i.e., sensor $k$ decides whether or not signal ${\cal A}$ is present. Let $d_k$ denote the binary decision of sensor $k$, where $d_k\!=\!0$ and $d_k\!=\!1$, respectively, correspond to $\mathcal{H}_{0}$ and $\mathcal{H}_{1}$. The test statistic for sensor $k$ is \vspace{-.12cm}
%------------------------------------------------------------------------------------------
\begin{equation}\label{energy_detection}
\Lambda_{k}=\frac{1}{N}\sum^{N}_{n=1}|x_{k,n}|^{2}   \gtrless
\begin{matrix}
{\scriptstyle d_k=1}\cr{\scriptstyle d_k=0}
\end{matrix} 
~~\theta_{k}
\vspace{-.12cm}
\end{equation}
%------------------------------------------------------------------------------------------
where $\theta_{k}$ is local decision threshold to be optimized. 
For the signal model in (\ref{xk}), conditioned on each hypothesis $x_{k}$ is Gaussian, that is, $x_{k}|\mathcal{H}_{0} \sim {\cal N} \left(0,\sigma_{w_{k}}^{2}\right)$ and $x_{k}|\mathcal{H}_{1} \sim {\cal N} \left({\cal A}\gamma_{g_{k}},\sigma_{w_{k}}^{2}\right)$.
The test statistic $\Lambda_{k}$ in (\ref{energy_detection}) has non-central Chi-square distribution [7] as given below \vspace{-.15cm}
%------------------------------------------------------------------------------------------
%\begin{equation}\label{Chi-square}
%\Lambda_{k} \sim
%\begin{cases}
%\chi^{2}_{N}(\eta_{k}) &~~~~~\mathcal{H}_1 \\
%\chi^{2}_{N} &~~~~~\mathcal{H}_{0}
%\end{cases} 
%\end{equation}
%
\begin{equation}\label{Chi-square}
\mathcal{H}_1: \Lambda_{k} \sim
\chi^{2}_{N}(\eta_{k}), ~~~\mathcal{H}_{0}: \Lambda_{k} \sim
\chi^{2}_{N} 
\end{equation}
%------------------------------------------------------------------------------------------
where $\eta_{k}\!=\!{\cal A}^2 \mathbb{E} \{ g^2_{k,n}\}\!=\!{\cal A}^{2} \gamma_{g_{k}}$ is 
%the received SNR value or 
the non-centrality parameter.
%------------------------------------------------------------------------------------------
%\begin{figure}[t]
		%\centering
		%\includegraphics[scale=.12]{Drawing2.pdf}
		%\caption{Model of an energy harvesting node}
		%\label{fig2}
%\end{figure}
%------------------------------------------------------------------------------------------
Using (\ref{Chi-square}), the false-alarm probability $P_{f_k}$ and detection probability $P_{d_k}$ can be derived as following \vspace{-.1cm}
%------------------------------------------------------------------------------------------
\begin{align}%\label{P_f}\label{P_Dk}
\vspace{-.12cm}
P_{f_k}= & \Pr (\Lambda_{k} >\theta_{k} |\mathcal{H}_{0} )= \frac{\Gamma\big(N/2,\frac{N\theta_{k}} {\sigma_{w_k}^{2}}\big) }{\Gamma\left(N/2\right)} \label{P_f} \\
P_{d_k}=& \Pr( \Lambda_{k}>\theta_{k} |\mathcal{H}_{1} ) = Q_{N/2}\big(\frac{\sqrt{\eta_{k}}}{\sigma_{w_{k}}},\frac{\sqrt {N\theta_{k}}}{\sigma_{w_{k}}}\big) \label{P_Dk}
\vspace{-.1cm}
\end{align}
%------------------------------------------------------------------------------------------
where $\Gamma(n)$ is the gamma function, $\Gamma(n,x)=\int^{\infty}_{x}t^{n-1}e^{-t}dt$ is the upper incomplete gamma function, $Q_n(a,b)=\int^{\infty}_{b}x(\frac{x}{a})^{n-1}\text{exp}(\frac{x^2+a^2}{-2})I_{n-1}(ax)dx$ is the generalized Marcum-Q function, and $ I_{n-1}(\cdot)$ is modified Bessel function of order $n-1$ \cite{8}.
%------------------------------------------------------------------------------------------
%
%In Fig. \ref{2}, model of an energy harvesting sensor is depicted. Each sensor has battery with capacity of ${\cal K}$ units, and it is able to harvest energy from the environment.

We assume each sensor is able to harvest energy from the environment and stores this harvested energy in a battery that has the capacity of storing at most ${\cal K}$ units of energy.
As shown in Fig. \ref{1}, the sensors communicate with the FC through orthogonal fading channels with channel gains $|h_k|$'s that are independent and have Rayleigh distribution with parameters $\gamma_{h_{k}}$.
The sensors employ on-off keying (OOK) signaling for communication, where a $d_k\!=\!1$ decision at sensor $k$ is conveyed at the cost of spending one or more energy units and a $d_k\!=\!0$ decision is conveyed through a no-transmission with no energy cost. We assume that only sending a message costs units of energy, and the energy of making the observation and processing is negligible. The number of energy units spent to convey a $d_k\!=\!1$ decision depends on the quality of the channel gain $|h_k|$ and the battery state of sensor $k$. Motivated by the channel-inversion power control strategy developed in the wireless communication community \cite{9}  we try to compensate for the fading and  let the number of energy units spent to convey a $d_k\!=\!1$ decision be (roughly) inversely proportional to $|h_k|$ (i.e., a smaller $|h_k|$ corresponds to a larger number of energy units), albeit if the battery has sufficient number of stored energy units. 
%
%Sensor $k$ consumes unites of energy to send positive message towards FC, where number of consuming units depends on the quality of the channel between sensor $k$ and the FC in a way that, the smaller the channel gain $|h_k|$ is (the worse the channel quality is), the more number of energy units is needed. Therefore, the sensor needs sufficient battery charge in order to send message in poor channel quality.
%
To avoid the battery depletion when $|h_k|$ is too small, we impose an extra constraint inspired by the channel truncation technique in the channel-inversion power control strategy \cite{9}, to ensure that a $d_k\!=\!1$ decision is conveyed only if $|h_k|$ exceeds a minimum threshold $\zeta_k$ (choice of $\zeta_k$ will be discussed later).
Let $t$ indicate the index of the observation period and $b_{k,t}$ denote the battery state of sensor $k$ in the observation period $t$.
Let $u_{k,t}$ represent the sensor output corresponding to the observation period $t$. Based on the above explanations, we define $u_{k,t}$ as
%
%------------------------------------------------------------------------------------------
\begin{equation}\label{u_k}
u_{k,t} =
\begin{cases}
\lceil \frac{\lambda }{|h_{k}|} \rceil  &~~~~~\Lambda_{k}>\theta_{k},~b_{k,t}>\lceil \frac{\lambda}{|h_{k}|} \rceil,~|h_{k}|^2>\zeta_{k} \\
0 &~~~~~~ \text{Otherwise}
\end{cases}
\vspace{-.12cm}
\end{equation}
%------------------------------------------------------------------------------------------
%where $h_k$ is the channel fading gain and we model each $K$ channel as flat
%Rayleigh fading where the fading coefficients have real values and we assume is perfect CSI and $|h_k|^2$ has exponential distribution with parameter $\gamma_{h_{k}}$.
%We assume $|h_k|$ has Rayleigh distribution, thus $|h_k|^2$ has exponential distribution with parameter $\gamma_{h_{k}}$.
where $\lambda$ is a power regulation constant (that depends on the battery structure). We use the round function $\lceil.\rceil$ toward $+\infty$, to ensure that $u_{k,t}$ is a discrete symbol and the energy of this symbol is equal to the number of consumed energy units to convey $d_k=1$. The constraint $\Lambda_{k}>\theta_{k}$ in (\ref{u_k}) comes directly from (\ref{energy_detection}).
%Note that $u_{k,t}$ is equal to  Number of units of energy required for transmission, is determined as $\lceil \frac{\lambda}{|h_{k}|} \rceil$. 
%\par Let $b_{k,t}$ denote the battery state of sensor $k$ in an observation period, $t$. 
%
We assume the average energy of the transmitted symbol $u_{k,t}$ is constrained, i.e., $P_{av_k}= \mathbb{E}\{\lceil \frac{\lambda }{|h_{k}|} \rceil^{2}\big|u_{k}=\lceil \frac{\lambda }{|h_{k}|} \rceil \}$, where the expectation is taken with respect to $|h_k|$.
%$\zeta_k$ in (\ref{u_k}) by setting a constraint on 
%Average power of transmitted signal is based on channel quality. More power is consumed in poor quality of channel. We will find optimal threshold based on average transmission power of sensor $k$ i.e. $P_{av_{k}}$. 
%
We model $b_{k,t}$ in (\ref{u_k}) as the following \vspace{-.15cm}
%------------------------------------------------------------------------------------------
\begin{equation}\label{b_k}
b_{k,t}=\text{min} \big\{b_{k,t-1}-\lceil \frac{\lambda}{|h_k|}\rceil I_{u_{k,t-1}} +\Omega_{k,t}~,~{\cal K} \big\}
\vspace{-.15cm}
\end{equation}
%------------------------------------------------------------------------------------------
where $b_{k,t-1}$ is the battery state of the previous observation period and $\Omega_{k,t} \in \{0,1\}$ is a binary random variable, indicating whether or not sensor $k$ harvests one unit of energy. 
%the arrival energy during time interval. We assume energy arrives in units and at each time interval the sensor is capable of harvesting at most one unit of energy. 
%
We assume $\Omega_{k,t}$ is
a Bernoulli random variable, with $\Pr (\Omega_{k,t} \!= \!1 ) \!=\! p_e$, where $p_e$ depends on the harvesting structure. This assumption is repeatedly used in the literature (see \cite{10} and references therein). The indicator function $I_{u_{k,t-1}}$ in (\ref{b_k}) is defined as
\vspace{-.3cm}
%------------------------------------------------------------------------------------------
\begin{equation} \label{I_k}
I_{u_{k,t-1}}=
\begin{cases}
1 &~~~~~u_{k,t-1}>0\\
0 &~~~~~~ \text{Otherwise}
\end{cases} 
\vspace{-.1cm}
\end{equation}
%------------------------------------------------------------------------------------------
In the remaining, we focus on one observation period and we drop the subscript $t$ from the battery state $b_{k,t}$ and the sensor output $u_{k,t}$.
Given our system model description above, our goal is to investigate the optimal local decision thresholds $\theta_k$'s  in (\ref{energy_detection}) that optimizes the detection performance metric.
%
%In order to save energy of limited battery capacity, we need to optimized system parameters. Our goal is to find optimal threshold $\theta_k$ for each sensor.
%------------------------------------------------------------------------------------------
\vspace{-1mm}
\section{Optimizing Local Decision Thresholds}\label{Section-3}
\vspace{-.12cm}
We consider two detection performance metrics to find the optimal $\theta_k$'s: (i) the detection probability at the FC, assuming that the FC utilizes the optimal fusion rule based on Neyman-Pearson optimality criterion, and (ii) the KL distance between the two distributions of the received signals at the FC conditioned on hypothesis $\mathcal{H}_{0}, \mathcal{H}_{1}$. In Section \ref{max-PD-fusion-rule} we derive the optimal fusion rule and the expressions for the detection and false alarm probabilities $P_D, P_F$ at the FC. In Section \ref{max-KL-FC} we derive two approximate expressions for the KL distance at the FC. In Section \ref{choose-zeta-threshold} we discuss the choice of the threshold $\zeta_k$ in (\ref{u_k}).
%In this section, first we will find threshold $\theta_k$ which maximizes Kullback-Leibler distance (KL) from sensors to the FC. Then, Under Neyman-Pearson (NP) test, we will  maximize probability at FC and find optimal $\theta_k$. In the simulation section, the we consider special case in which all the sensors have same local threshold, i.e., $\theta_k=\theta$.
%
\vspace{-2.5mm}
\subsection{Optimal LRT Fusion Rule and $P_D, P_F$ Expressions}\label{max-PD-fusion-rule}
\vspace{-1mm}
The received signal at the FC from sensor $k$ is $y_{k}=h_{k}u_{k}+n_{k}$, where the additive communication channel noise $n_{k}\sim {\cal N}\left(0,\sigma^{2}_{n_{k}}\right)$. 
The likelihood ratio  at the FC is \cite{11}
\vspace{-1mm}
%------------------------------------------------------------------------------------------
\begin{eqnarray}\label{LRT}
\Delta_{\text{LRT}}&=&\log \left(\frac{f\left(y_{1}, ..., y_K|\mathcal{H}_{1}\right)}{f\left(y_{1},..., y_K| \mathcal{H}_{0}\right)}\right)\nonumber\\
&=&\sum_{k=1}^K \log\left(\frac{\sum_{u_{k}}f\left(y_{k}|u_k,\mathcal{H}_{1}\right)\Pr\left(u_k|\mathcal{H}_{1}\right)}{\sum_{u_{k}}f\left(y_{k}|u_k,\mathcal{H}_{0}\right)\Pr\left(u_k|\mathcal{H}_{0}\right)}\right)
\vspace{-2mm}
\end{eqnarray}
%------------------------------------------------------------------------------------------
in which we use the fact that, given $\mathcal{H}_{i}$ the received signals at the FC are independent, i.e., $f(y_1,..., y_K|\mathcal{H}_{i})=\prod_{k=1}^K f(y_k|\mathcal{H}_{i})$.
Examining (\ref{LRT}), we note given $u_k$, $y_k$ and $\mathcal{H}_{i}$ are independent and hence $f\left(y_{k}|u_k,\mathcal{H}_{i}\right)\!=\!f\left(y_{k}|u_k\right)$ for $i\!=\!0,1$. Also, given $u_k$, $y_k$ is Gaussian, i.e., $y_{k}|_{u_{k}=0}\sim {\cal N}\left(0,\sigma^{2}_{n_{k}}\right)$ and $y_{k}|_{u_{k}=\lceil \frac{\lambda }{|h_{k}|} \rceil}\sim {\cal N}\left(\lceil \frac{\lambda }{|h_{k}|} \rceil h_{k},\sigma^{2}_{n_{k}}\right)$.
The probabilities $\Pr( u_{k}|\mathcal{H}_{1})$, $\Pr( u_{k}|\mathcal{H}_{0})$ in (\ref{LRT}) are
\vspace{-0.2cm}
%------------------------------------------------------------------------------------------
\begin{eqnarray}
&& \Pr \big (u_{k}=\lceil \frac{\lambda }{|h_{k}|} \rceil \big | \mathcal{H}_{1} \big ) \nonumber\\
&= & \Pr \Big(\Lambda_{k}\!>\!\theta_{k} , ~ b_{k}\!>\!\lceil \frac{\lambda}{|h_{k}|} \rceil , ~|h_{k}|^2\!>\!\zeta_{k} \big| \mathcal{H}_{1} \Big ) \nonumber \\ 
&= & \Pr \big (\Lambda_{k}>\theta_{k} | \mathcal{H}_{1} \big ) \Pr \big( b_{k}>\lceil \frac{\lambda}{|h_{k}|} \rceil \big ) \Pr \big (|h_{k}|^2>\zeta_{k} \big ) \nonumber \\ 
&= & P_{d_k}\rho_{k}q_{k}=\alpha_{k} \label{alpha_k}\\
&& \Pr \big (u_{k}=\lceil \frac{\lambda }{|h_{k}|} \rceil |\mathcal{H}_{0} \big) \nonumber\\
&= & \Pr \big(\Lambda_{k}>\theta_{k} \big| \mathcal{H}_{0}\big) \Pr\big( b_{k}>\lceil \frac{\lambda}{|h_{k}|} \rceil \big) \Pr\big (|h_{k}|^2>\zeta_{k}\big) \nonumber \\ 
&= & P_{f_k}\rho_{k}q_{k}=\beta_{k}\label{beta_k}
\vspace{-4mm}
\end{eqnarray}
%------------------------------------------------------------------------------------------
%\begin{figure}[t]
	%\centering
	%\includegraphics[scale=.5]{CDF_Pe_k20.eps}
	%\caption{CDF of $b_k$ for ${\cal K} = 20 $ and $p_e=0.5, 0.75, 0.82 $}
	%\label{fig3}
%\end{figure}
%------------------------------------------------------------------------------------------
%\begin{figure}[t]
	%\centering
	%\includegraphics[scale=.5]{PDF_k_50.eps}
	%\caption{PMF of $b_k$ for ${\cal K} = 50 $ and $p_e=0.8$}
	%\label{fig4}
%\end{figure}
%----------------------------------------------------------------------------------
\begin{figure}
   
    \subfigure[]
    {
        \includegraphics[scale=.277]{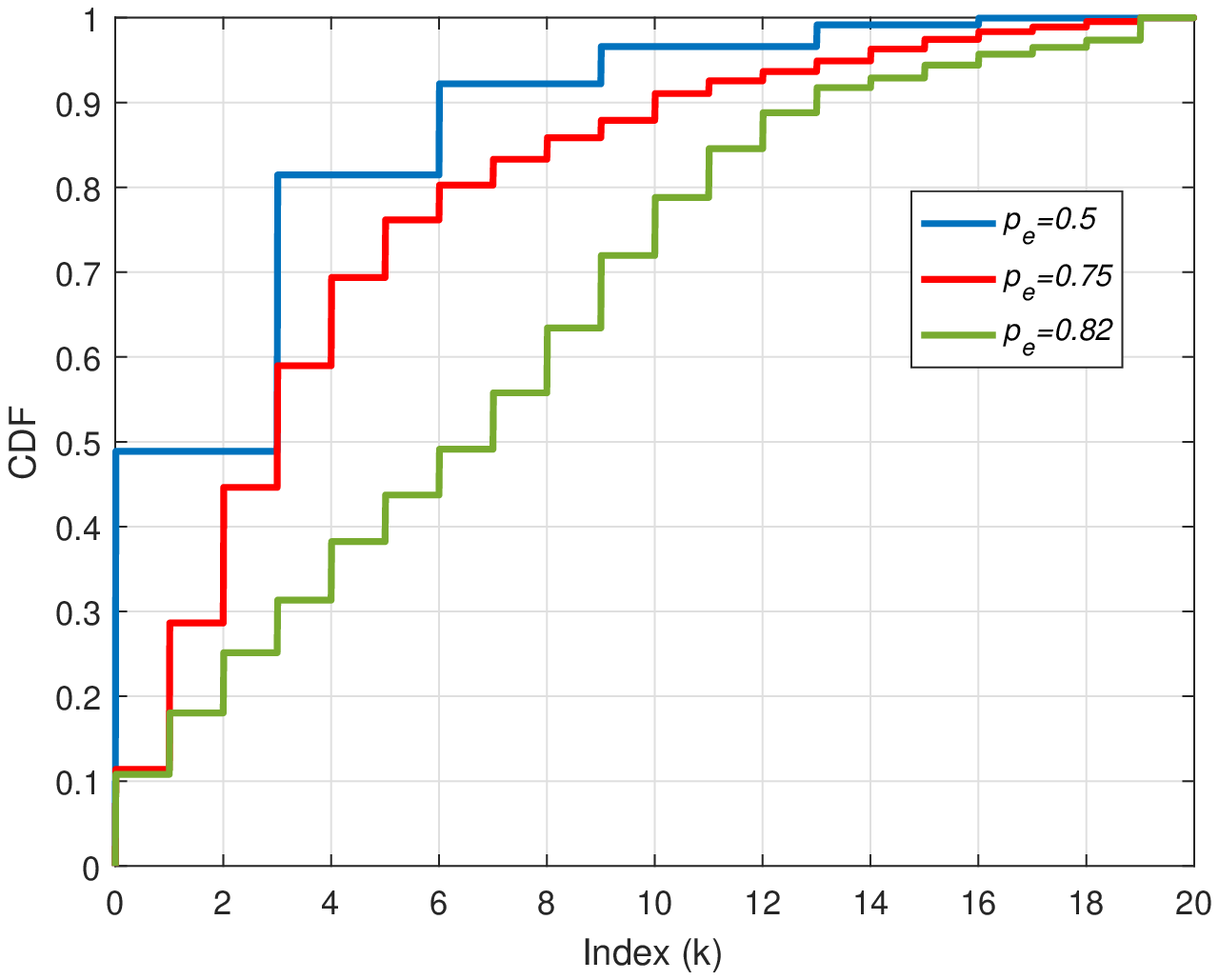}\label{fig2a}
    }
    \subfigure[]
    {
        \includegraphics[scale=.277]{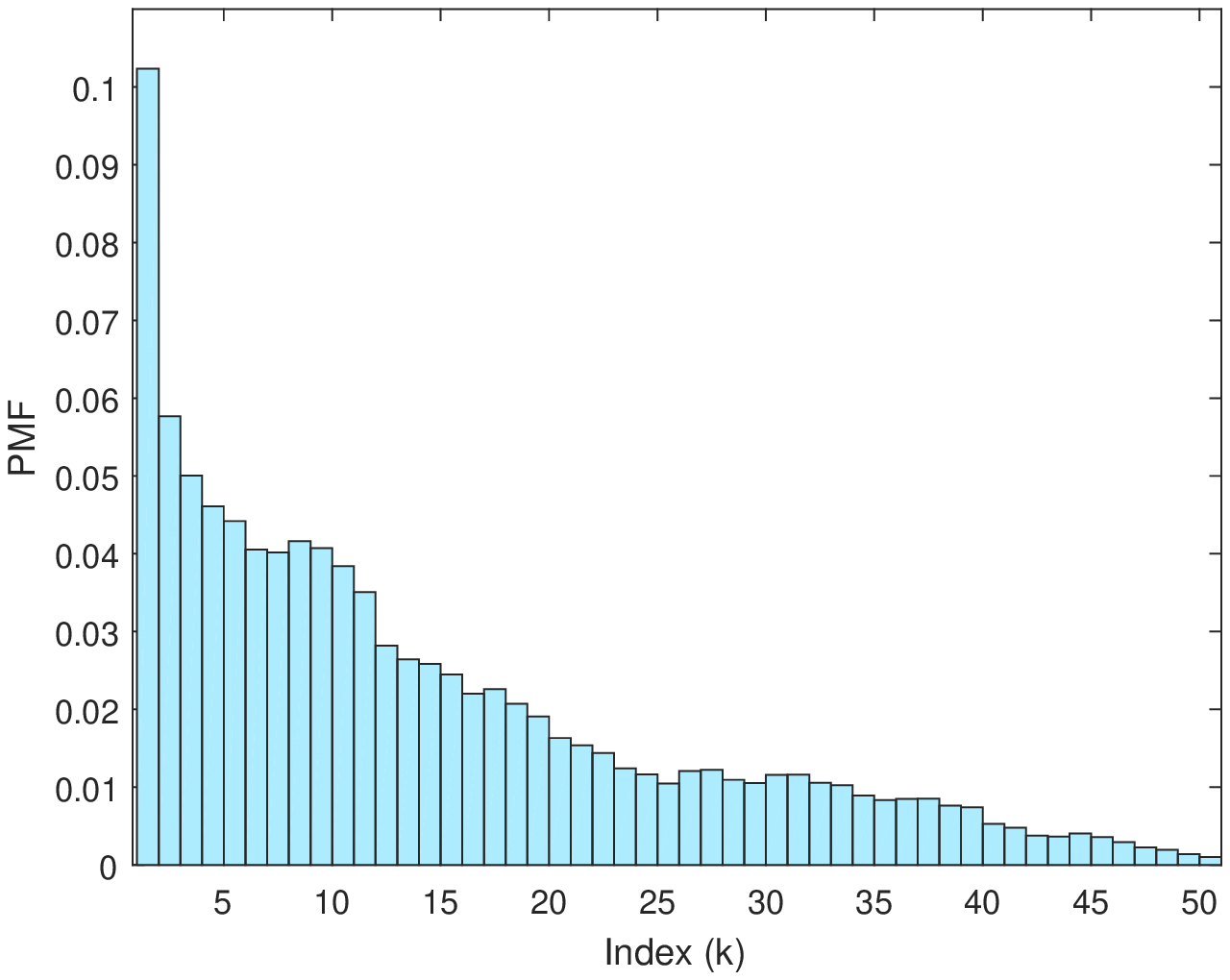}\label{fig2b}
    }
    \caption
    {
        (a) CDF of $b_k$ for ${\cal K} \!= \! 20 $ and $p_e\!=\!0.5, 0.75, 0.82 $, 
        (b) pmf of $b_k$ for ${\cal K} \!= \!50 $ and $p_e\!=\!0.8$.
    }
    \label{fig2}
    \vspace{-.6cm}
\end{figure}
%----------------------------------------------------------------------------------
%------------------------------------------------------------------------------------------
%\vspace{-0.3cm}
%\begin{eqnarray}\label{beta_k}
%\Pr ( u_{k}=\lceil \frac{\lambda }{|h_{k}|} \rceil |\mathcal{H}_{0} ) =P_{fk}(1-p_{0})q_{k}=\beta_{k}
%\end{eqnarray}
%------------------------------------------------------------------------------------------
\noindent where $P_{f_k}, P_{d_k}$ are given in (\ref{P_f}), (\ref{P_Dk}),  $\rho_k\!=\!\Pr (b_{k}>\lceil \frac{\lambda}{|h_{k}|} \rceil)$ 
and $q_k\!=\!\Pr (|h_{k}|^2>\zeta_{k})\!=\!\text{exp}(-\zeta_k/\gamma_{h_{k}})$.
%We use numerical evaluations to find $\rho_{k}$. 
Assuming $b_k$ in (\ref{b_k}) is a stationary random process, one can compute the cumulative distribution function (CDF) and the probability mass function (pmf) of $b_k$ in terms of ${\cal K}, p_e, \gamma_{h_k}$.
%by conducting some simulation, we calculate the CDF and PMF of $b_k$ when the battery is in its steady state. 
Fig.\ref{fig2a} shows CDF of $b_k$ for ${\cal K}\!=\!20 $ and $p_e\!=\!0.5, 0.75, 0.82$, and Fig.\ref{fig2b} depicts pmf of $b_k$ for  ${\cal K}\!=\!50 $ and $p_e\!=\!0.8$. For our numerical results in Section \ref{last-section} we use pmf of $b_k$ to find $\rho_{k}$ in (\ref{alpha_k}) and (\ref{beta_k}).
%------------------------------------------------------------------------------------------
Combing all, we can rewrite $\Delta_{\text{LRT}}$ as the following \cite{12}
%------------------------------------------------------------------------------------------
\vspace{-.1cm}
\begin{align*}\label{LRT2}
\Delta_{\text{LRT}}
\! =\! &\sum_{k=1}^K \! \log\!  \left(\frac{\alpha_{k}f(y_{k}|u_{k}\! =\! \lceil \frac{\lambda }{|h_{k}|} \rceil)+(1\! -\! \alpha_{k})f(y_{k}|u_{k}\! =\!  0)}{\beta_{k}f(y_{k}|u_{k}\! =\! \lceil \frac{\lambda }{|h_{k}|} \rceil)+(1\! -\! \beta_{k})f(y_{k}|u_{k}\! =\! 0)}\! \right) \nonumber\\ 
= &\sum_{k=1}^{K}\!\log\!\frac{\alpha_{k}\text{exp}\big({-\frac{(y_{k}\!-\!\lceil \frac{\lambda }{|h_{k}|} \rceil h_{k})^{2}}{2\sigma^{2}_{n_{k}}}}\big)\!+\!(1-\alpha_{k})\text{exp}\big({-\frac{y_{k}^{2}}{2\sigma^{2}_{n_{k}}}}\big)}{\beta_{k}\text{exp}\big ({-\frac{(y_{k}\!-\!\lceil \frac{\lambda }{|h_{k}|} \rceil h_{k})^{2}}{2\sigma^{2}_{n_{k}}}}\big )\!+\!(1-\beta_{k})\text{exp}\big ({-\frac{y_{k}^{2}}{2\sigma^{2}_{n_{k}}}}\big )}
\vspace{-.8cm}
\end{align*}
%------------------------------------------------------------------------------------------
In low SNR regime as $\sigma^{2}_{n_{k}}\!\rightarrow \!\infty$ taking a logarithm from $\Delta_{\text{LRT}}$ and using the
approximations $e^{-x} \approx1-x$ and $\log(1+x)$ for small
$x$, we can simplify $\Delta_{\text{LRT}}$ to $\Delta_{\text{LRT}}\! \approx \!-T_{k}+\sum_{k=1}^{K}\nu_{k}y_{k}$ where $T_{k}\!=\!\sum_{k=1}^{K}\lceil \frac{\lambda }{|h_{k}|} \rceil^{2} h_{k}^{2}(\alpha_{k}-\beta_{k})/2\sigma^{2}_{n_{k}}$ and  $\nu_{k}\!=\!\lceil \frac{\lambda }{|h_{k}|} \rceil h_{k}(\alpha_{k}-\beta_{k})/\sigma^{2}_{n_{k}}$.
%------------------------------------------------------------------------------------------
Given a threshold $\tau$, the optimal likelihood ratio test (LRT) is $\Delta_{\text{LRT}}  \gtrless
\begin{matrix}
{\scriptstyle \mathcal{H}_{1}}\cr{\scriptstyle \mathcal{H}_{0}}
\end{matrix} 
\tau$.
The false alarm and detection probabilities $P_{F},P_D$ at the FC are
%------------------------------------------------------------------------------------------
\begin{eqnarray}
P_{F}&=&\Pr\left( \Delta_{\text{LRT}}>\tau|\mathcal{H}_{0}\right)=Q\big(\frac{\tau-\mu_{\Delta|\mathcal{H}_{0}}}{\sigma_{\Delta|\mathcal{H}_{0}}}\big) \label{PF-at-FC}\\
P_{D}&=&\Pr\left(\Delta_{\text{LRT}}>\tau|\mathcal{H}_{1}\right)\nonumber \\ 
&= & Q\left(\frac{Q^{-1}(a)\sigma_{\Delta|\mathcal{H}_{0}}+\mu_{\Delta|\mathcal{H}_{0}}-\mu_{\Delta|\mathcal{H}_{1}}}{\sigma_{\Delta|\mathcal{H}_{1}}}\right) \label{PD-at-FC}
\end{eqnarray}
%------------------------------------------------------------------------------------------
where 
\vspace{-.2cm}
\begin{eqnarray}\label{parameters}
\mu_{\Delta|\mathcal{H}_{i}}\!=\! -T_{k}\!+\!\sum_{k=1}^{K}\nu_{k}\mu_{y_{k}|\mathcal{H}_{i}}, ~~
\sigma^{2}_{\Delta|\mathcal{H}_{i}}\!= \! \sum_{k=1}^ {K}\nu_{k}^{2} \sigma_{y_{k}|\mathcal{H}_{i}}^{2}, i=0,1 \label{PD in FC-para} \nonumber\\
\mu_{y_{k}|\mathcal{H}_{0}}=\lceil \frac{\lambda }{|h_{k}|} \rceil h_{k} \beta_{k}\nonumber,~~ \sigma_{y_{k}|\mathcal{H}_{0}}^{2}\!=\!\lceil \frac{\lambda }{|h_{k}|} \rceil^{2} h_{k}^{2}  \beta_{k}(1\!-\!\beta_{k})\!+\!\sigma^{2}_{n_{k}}\\ \nonumber
\mu_{y_{k}|\mathcal{H}_{1}}\!=\!\lceil \frac{\lambda }{|h_{k}|} \rceil h_{k} \alpha_{k},~~ \sigma_{y_{k}|\mathcal{H}_{1}}^{2}\!=\!\lceil \frac{\lambda }{|h_{k}|} \rceil^{2} h_{k}^{2}  \alpha_{k}(1\!-\!\alpha_{k})\!+\!\sigma^{2}_{n_{k}} \nonumber
\end{eqnarray}
%
%The function $Q(x)=\int^{\infty}_{x}\frac{1}{\sqrt{2\pi}}e^{-\frac{t^{2}}{2}}dt$. 
The threshold $\tau$ is determined from the constraint on $P_F \! \leq \! a $ in terms of $a$. 
%
%Assuming $P_{F} = a$ as the NP constraint, $\tau$ can be derived in terms of $a$ and then detection probability $P_{D}$ is written as:
%------------------------------------------------------------------------------------------
%
%------------------------------------------------------------------------------------------
%Therefore, $P_D$ in (\ref{eq22}) can be written as:
%------------------------------------------------------------------------------------------
%\begin{equation*}\label{eq24}
%\resizebox{1\hsize}{!}{$ \displaystyle{ P_{D}=Q\!\left(\frac{Q^{-1}(a)\sqrt{\sum^{M}_{k=1}\nu^{2}_{k}\sigma_{y_{k}|\mathcal{H}_{0}}^{2}}+\sum^{M}_{k=1}\!\nu_{k}\lceil \frac{\lambda }{|h_{k}|} \rceil h_{k} \left(\beta_{k}-\alpha_{k}\right)}{\sqrt{\sum^{M}_{k=1}\nu^{2}_{k}\sigma_{y_{k}|\mathcal{H}_{1}}^{2}}}\right)}$}
%\end{equation*}
%------------------------------------------------------------------------------------------
We note that $P_D$ expression depends on all our optimization variables $\theta_k$'s through $\alpha_k,\beta_k$'s in $\mu_{\Delta|\mathcal{H}_{i}}$ and $\sigma^{2}_{\Delta|\mathcal{H}_{i}}$. 
%%%%%%%%%%%%%%%%%%%%%%%%%%%%%%%%%%%%%%%%%%%%%%%%%%%%%%%%%%%%%%%%%%%%%%%%%%%%%%%%%%%%%%%%%%%
\subsection{$KL$ Expression}\label{max-KL-FC}
 \vspace{-1mm}
%------------------------------------------------------------------------------------------
\begin{figure}[t]
	\centering
	\includegraphics[scale=.35]{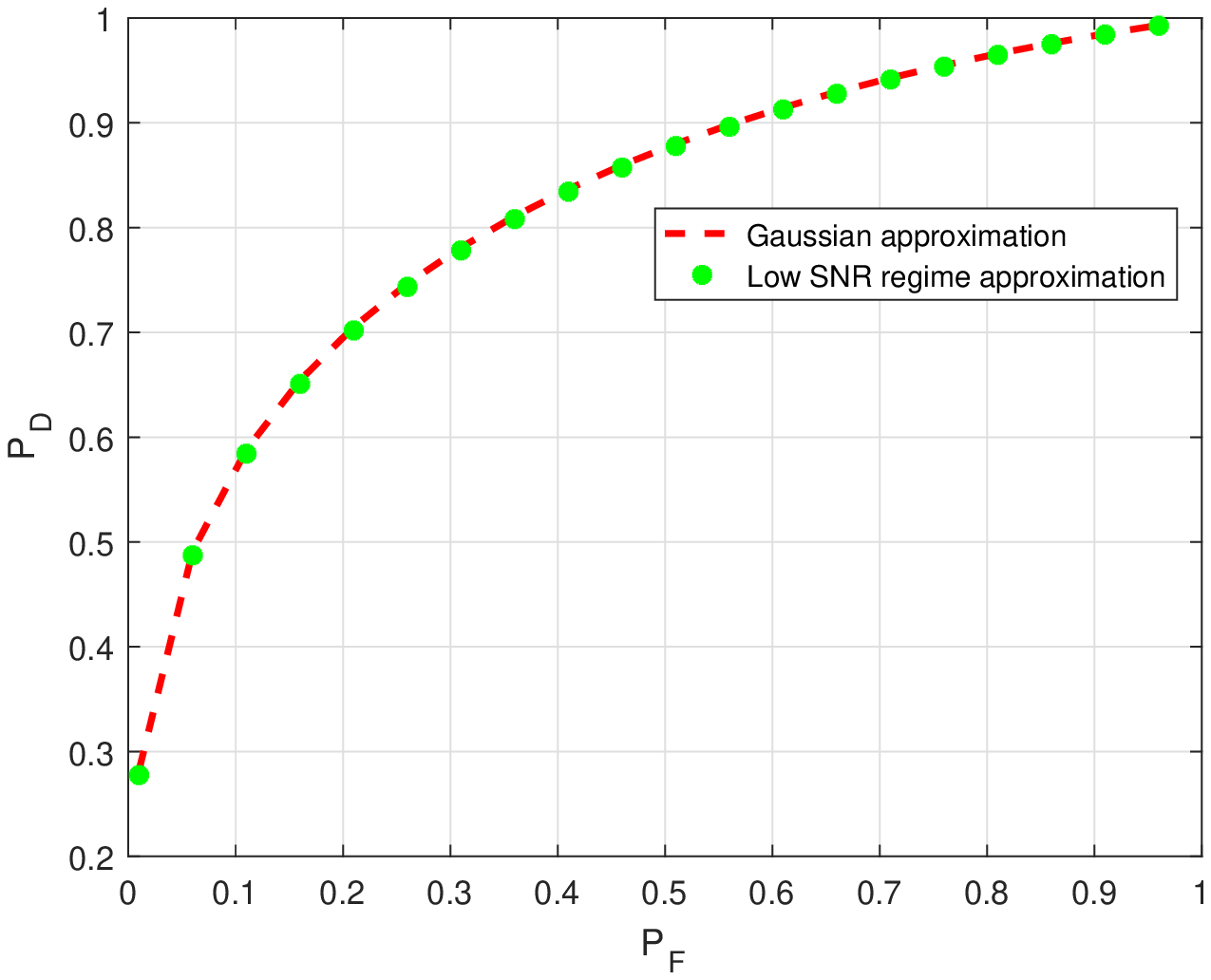}
	\caption{$P_D$ vs. $P_F$, ${\cal K}=20, p_e=0.75, P_{av}=1$dB.}
	\label{fig4}
    \vspace{-5mm}
\end{figure}
%----------------------------------------------------------------------------------------
Let $KL_{tot}$ denote the KL distance between the two distributions $f(y_1,...,y_K|\mathcal{H}_{1})$ and $f(y_1,..., y_K|\mathcal{H}_{0})$ at the FC. Since $f(y_1,..., y_K|\mathcal{H}_{i})\!=\!\prod_{k=1}^K f(y_k|\mathcal{H}_{i})$, we have $KL_{tot}\!=\!\sum_{k=1}^K KL_k$ where $KL_k$ by definition is \cite{13}
\begin{equation}\label{kl}
KL_k=\int_{y_k} f(y_{k}|\mathcal{H}_{1})\log \left (\frac{f(y_{k}|\mathcal{H}_{1})}{f(y_{k}|\mathcal{H}_{0})}\right)dy_k
\vspace{-3mm}
\end{equation}
%------------------------------------------------------------------------------------------
%Knowing ${\sum_{u_{k}}f\left(y_{k}|u_k,\mathcal{H}_{i}\right)\Pr\left(u_k|\mathcal{H}_{i}\right)}\big)$ for $i=0,1$ and probabilities in (\ref{alpha_k}) and (\ref{beta_k})
%------------------------------------------------------------------------------------------
We note that the distributions $f(y_k|H_i), i\!=\!0,1$ are Gaussian mixtures and thus $KL_k$ in (\ref{kl}) does not have a general closed-form expression \cite{14} and approximations must be made. One can approximate $KL_k$ in (\ref{kl}) by the KL distance of two Gaussian distributions with the means $\mu_{y_{k}|\mathcal{H}_{0}}$, $\mu_{y_{k}|\mathcal{H}_{1}}$, and the variances $\sigma_{y_{k}|\mathcal{H}_{0}}^{2}$ and  $\sigma_{y_{k}|\mathcal{H}_{1}}^{2}$, respectively, i.e., $KL_k$ can be approximated as \cite{15}
%Remember $f(y_k| u_k, {\cal H}_i)$ for $i=0,1$ is Gaussian and $f(y_k|H_i)$ is a linear combination of Gaussian, so, it is a Gaussian mixture. By approximating $f(y_k|H_i)$ as Gaussian, 
%we can write (\ref{kl}) as \cite{16}
\begin{equation}\label{kl2}
\resizebox{1\hsize}{!}{$\displaystyle{KL_{k} \approx \frac{1}{2}\log (\frac{\sigma_{y_{k}|\mathcal{H}_{0}}^{2}}{\sigma_{y_{k}|\mathcal{H}_{1}}^{2}})+\frac{\sigma_{y_{k}|\mathcal{H}_{1}}^{2}  -  \sigma_{y_{k}|\mathcal{H}_{0}}^{2} + (\mu_{y_{k}|\mathcal{H}_{1}} - \mu_{y_{k}|\mathcal{H}_{0}})^{2}}{2\sigma_{y_{k}|\mathcal{H}_{0}}^{2}}}$}
\end{equation}
%--------------------------------------------------------------------------------------------
%By replacing parameters of (\ref{parameters}) in (\ref{kl2}), we can find a closed-form expression for the $KL_k$ of $\theta_k$. 
%As you can see, (\ref{kl2}) can not be more simplified, therefore in simulation section, we numerically find the $\theta_k$ which maximize the corresponding $KL$. 
%Spacial case\!:\! since we consider inhomogeneous sensor network and  the observations at the sensors, conditioned on the true hypothesis, are independent, the total $KL$ at FC simplifies greatly and decouples, i.e., $KL_\text{Total}=\sum_{k=1}^{M}KL_k$. 
%In simulation section we will compare performance of thresholds which obtained from maximizing $KL_k$ and $KL_\text{Total}$.
Another approximation for $KL_{k}$ in (\ref{kl}) can be found using the low SNR regime approximation in Section \ref{max-PD-fusion-rule}, as the following
%
%By using low SNR regime approximation, we can write (\ref{kl}) as \vspace{-.5cm}
%----------------------------------------------------------------------
\begin{align} \label{kl3}
\vspace{-.4cm}
\resizebox{.9\hsize}{!}{$ KL_k\!\approx\!c_k(\beta_k\!-\!\alpha_k)\bigg\{c_k\sqrt{\frac{\pi}{2\sigma^2_{n_k}}}\Big((1-\alpha_k)(Q(\frac{y_k}{\sigma_{n_k}})-0.5) $}\nonumber \\
\resizebox{1\hsize}{!}{$+\alpha_k Q(\frac{y_k\!-\!c_k}{\sigma_{n_k}})\Big)\!+\! \alpha_k\exp \big(\frac{(c_k\!-\!y_k)^2}{-2\sigma^2_{n_k}}\big)+(1\!-\!\alpha_k)\exp \big(\frac{-y_k^2}{2\sigma^2_{n_k}}\big)\bigg\}$} 
\vspace{-2cm}
\end{align}
where $c_k=\lceil \frac{\lambda }{|h_{k}|} \rceil h_{k}$. Different from $P_D$ expression that depends on all $\theta_k$'s, $KL_{tot}$ is decoupled such that $KL_k$ depends on $\theta_k$ only through $\alpha_k,\beta_k$'s in $\mu_{y_k|\mathcal{H}_{i}}$ and $\sigma^{2}_{y_k|\mathcal{H}_{i}}$. 
%------------------------------------------------------------------------------------------
%\par $\zeta_{k}$ is the threshold of channel quality. 
%As we mention before, if the signal quality is greater than threshold $\theta_k$, the channel quality is greater than threshold $\zeta_{k}$ and the battery has sufficient charge, the sensor can send its message to FC. Otherwise it remains silent. 
\vspace{-.2cm}
\subsection{Choosing Threshold $\zeta_k$ in (\ref{u_k})}\label{choose-zeta-threshold}
We find $\zeta_k$ in (\ref{u_k}) via solving the constraint $P_{av_k}\!=\!\mathbb{E}\{\lceil \frac{\lambda }{|h_{k}|} \rceil^{2}\big|u_{k}=\lceil \frac{\lambda }{|h_{k}|} \rceil \}$. 
%Average power of transmitted signal is based on channel quality. More power is consumed in poor quality of channel. We will find optimal threshold based on average transmission power of sensor $k$ i.e. $P_{av_{k}}$. 
%Assume $P_{av_k}=E\{\lceil \frac{\lambda }{|h_{k}|} \rceil^{2}\}$, where the expectation is taken with respect to $|h_k|$. 
Recall $h_k$ has Rayleigh distribution. After some algebraic manipulations we obtain
%------------------------------------------------------------------------------------------
\vspace{-.15cm}
\begin{equation}\label{P_av}
\resizebox{1 \hsize}{!}{$ \displaystyle{P_{av_k}\!=\!\alpha_{k}\!\sum^{\infty}_{i=1}\!\left(i+1\right) \! \left (e^{\frac{-1}{\gamma_{h_{k}}}\text{max} \big\{\zeta_{k},\frac{\lambda^{2}}{i+1}\big \} }\!-\!e^{\frac{-\lambda^{2}}{i\gamma_{h_{k}}}}\right)u\big[\frac{\lambda^{2}}{i}\!- \! \zeta_{k}\big]}$}
\end{equation}
%------------------------------------------------------------------------------------------
where $u[.]$ is the step function and $\alpha_k$ is given in (\ref{alpha_k}). Note $\alpha_k$ depends on $\zeta_k$ through $q_k$. Although there is no explicit expression for $\zeta_k$, for our numerical results in Section \ref{last-section} we use (\ref{P_av}) to find $\zeta_{k}$ given $P_{av_k}$ via the interpolation technique. 
%As you can see from (\ref{P_av}), it is not straightforward to write $\zeta_{k}$ based on $P_{av_k}$. 
%Therefore in simulations we use interpolation to find $\zeta_{k}$ based on given $P_{av_k}$. 
%%%%%%%%%%%%%%%%%%%%%%%%%%%%%%%%%%%%%%%%%%%%%%%%%%%%%%%%%%%%%%%%%%%%%%%%%%%%%%%%%%%%%%%%%%%
\begin{figure}[t]
	\centering
	\includegraphics[scale=.4]{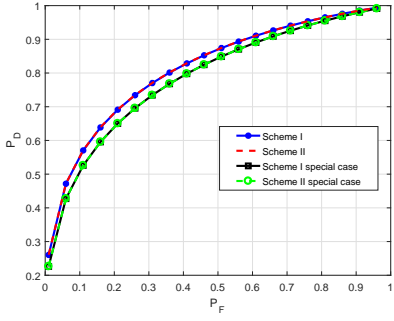}
	\caption{$P_D$ vs. $P_F$, ${\cal K}=20, p_e=0.75, P_{av}=1$ dB.}
	\label{fig3-a}
    \vspace{-4mm}
\end{figure}
%--------------------------------------------------------------------
\begin{figure}[t]
	\centering
	\includegraphics[scale=.4]{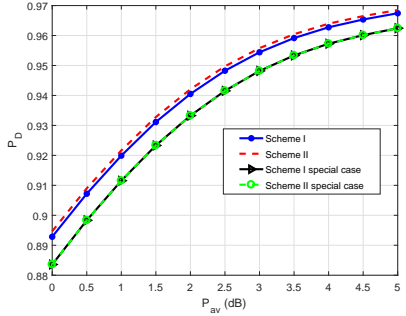}
	\caption{$P_D$ vs. $P_{av}$, ${\cal K}=20, p_e=0.75, P_F=0.5$.}
	\label{fig3-b}
    \vspace{-4mm}
\end{figure}
%--------------------------------------------------------------------
\begin{figure}[t]
	\centering
	\includegraphics[scale=.42]{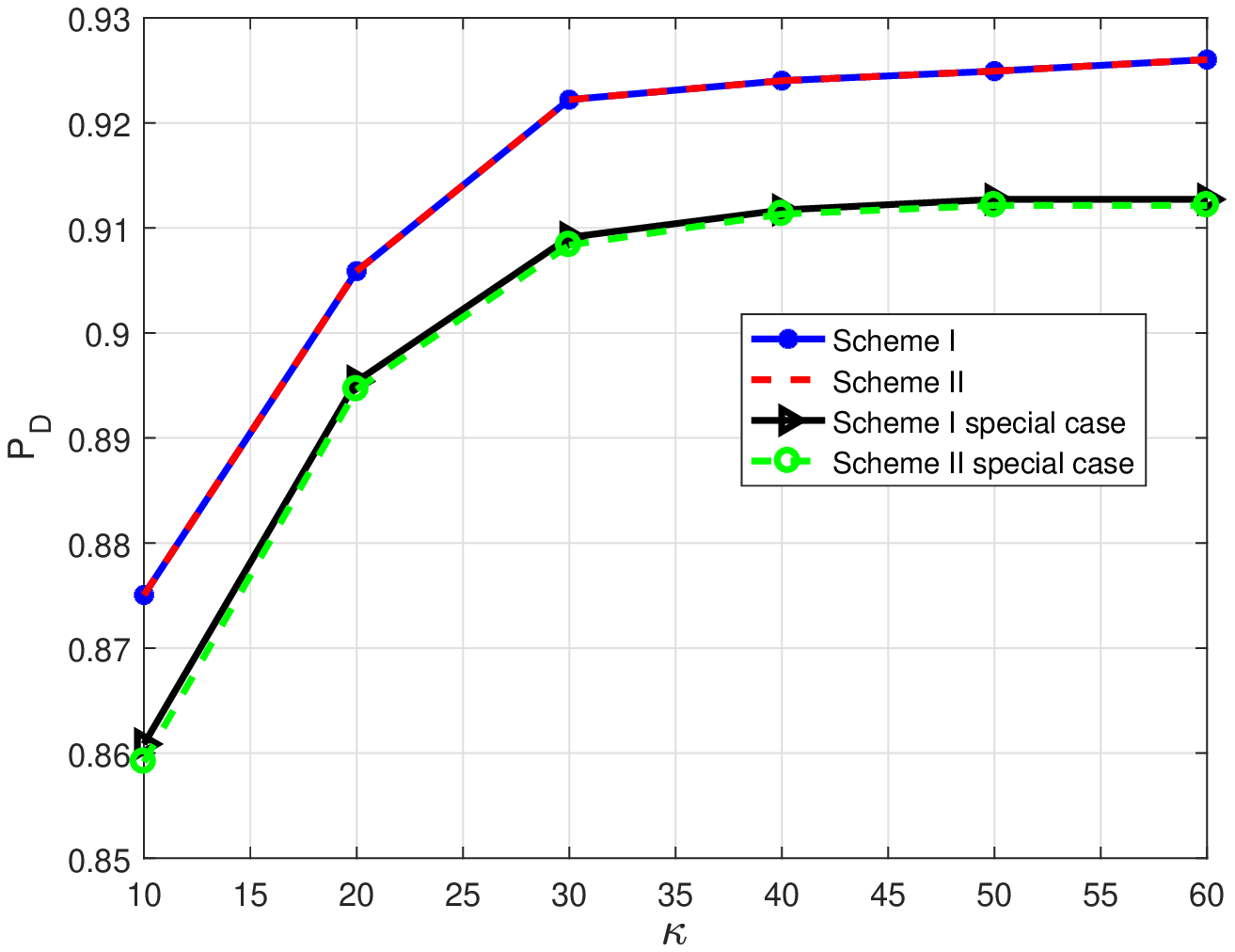}
	\caption{$P_D$ vs. ${\cal K}$, $p_e=0.8, P_{{av}}=1~\text{dB}, P_F=0.5$.}
	\label{fig3-c}
    \vspace{-5mm}
\end{figure}
%--------------------------------------------------------------------
\vspace{-.4cm}
\section{Simulation results and Conclusions}\label{last-section}
\vspace{-.1cm}
In this section, we numerically (i) find $\theta_k$'s which maximize $P_D$ in (\ref{PD-at-FC}). Finding   $\theta_k$'s in this case requires $K$-dimensional search, as $K$ grows the computational complexity grows exponentially; (ii) $\theta_k$'s which maximize $KL_{tot}=\sum_{k=1}^K KL_k$, using the $KL_k$ approximations in (\ref{kl2}), (\ref{kl3}). Finding $\theta_k$ in this case requires only one dimensional search and is computationally very efficient.  
We then compare $P_D$ evaluated at the $\theta_k$'s obtained from maximizing $P_D$ (refer to as scheme I in the plots), with $P_D$ evaluated at the $\theta_k$'s obtained from maximizing $KL_{tot}$ (refer to as scheme II in the plots). 
%
%and compare the result with the special case which we maximize $P_D$ considering same local threshold for all sensors.
%
%provide simulation results are provided to demonstrate the effectiveness of our proposed schemes. 
%
Our simulation parameters are  $K\!=\!3$,  ${\cal A}\!=\!1$, $N\!=\!100$, $\lambda\!=\!1$, $\gamma_h=[1.5,0.8,1.4]$, $\gamma_g=[1.3,2,0.9]$ and $\sigma_n^2=[0.9,1.2,0.8]$. Note that sensors are heterogeneous, in the sense that their statistical information parameters are different.
Given $P_{av_{k}}\!=\!P_{av}$ we first obtain numerically $\zeta_k$'s  using (\ref{P_av}), where $\zeta_k$'s are still different since $\alpha_k$'s are different.

%For acquiring $\theta_k$ for each sensor, four schemes are considered. In the first scheme, by maximizing $KL_k$ in (\ref{kl}) for each sensor,  optimal $\theta_k$s are obtained. In the second scheme, by $3$-dimensional search, we find $\theta_k$s which maximize $P_D$ at FC. 
%Other two schemes are special cases of previous ones i.e. We consider the optimal $\theta$ for all sensors. In the third scheme, the optimal $\theta$ is obtained by maximizing $KL_{\text{Total}}$. And, the last scheme, by $1$-dimensional search, the optimal $\theta$ is found which maximize $P_D$ at FC. 
%%
% 
%
%we compare the performance of two approximation mentioned before. 
%
Fig. \ref{fig4} plots $P_D$ versus $P_F$, where for each $P_F$ we evaluate $P_D$ using $\theta_k$'s which maximize $KL_{tot}$, based on the $KL_k$ approximations in (\ref{kl2}) and (\ref{kl3}). The fixed parameters in Fig. (\ref{fig4}) are ${\cal K}\!=\!20$ units, $p_e\!=\!0.75$, $P_{av}\!=\!1$ dB. 
%
%using $\theta_k$'s which maximize $KL_{tot} = \sum^K_{k=1} KL_k$, in the $KL_k$ approximation by (\ref{kl2}) and (\ref{kl3}). 
%
This figure shows that, these two approximations have similar $P_D-P_F$ behavior. Therefore, in the remaining figures, we use the $KL_k$ approximation in (\ref{kl2}).

Fig. \ref{fig3-a} depicts $P_D$ versus $P_F$ for ${\cal K}\!=\!20$ units, $p_e\!=\!0.75$, $P_{av}\!=\!1$ dB. To plot Fig. \ref{fig3-a}, for each $P_F$ we evaluate $P_D$ using $\theta_k$'s that maximize $P_D$ (scheme I) and $KL_{tot}$ (scheme II).
%By using optimal  $\theta_k$s of four schemes and (\ref{PD-at-FC}), we calculate $P_D$ for different amount of $P_F$.
%
Comparing schemes I and II in Fig. \ref{fig3-a}, we observe that these schemes perform very closely, indicating that using $\theta_k$'s that are obtained from maximizing $KL_{tot}$ are near-optimal.  
In Fig. \ref{fig3-a}, we also compare schemes I and II for the special case where we assume all sensors employ the same local threshold $\theta_k\!=\!\theta$. For this special case, finding $\theta$ maximizing $P_D$ or $KL_{tot}$ only needs one dimensional search. The performance gap between each scheme and its corresponding special case indicates that when sensors are heterogeneous, it is advantageous to use different local thresholds according to sensors' statistics (i.e., $\gamma_{h_k},\gamma_{g_k}, \sigma_{n_k}$).  
%
%considering different $\theta_k$ has better performance in inhomogeneous networks. 
%
%
%So, it is more practical to find $\theta_k$s by maximizing $KL_k$ which needs one dimensional search instead of applying $K$-dimensional search and maximizing $P_D$ at FC.

Fig. \ref{fig3-b} plots $P_D$ versus $P_{av}$ for ${\cal K}\!=\!20$ units, $p_e\!=\!0.75$, $P_F\!=\!0.5$. %$P_{av}$ varies from $0$ to $5dB$. 
As expected, $P_D$ increases as $P_{av}$ increases. The reason is as $P_{av}$ increases $\zeta_k$'s decrease, and sensors can afford to transmit even when their channel gains are weaker.
%, sensor $k$ can transmit its data towards FC in weaker channel. 
%

Fig. \ref{fig3-c} illustrates $P_D$ versus ${\cal K}$ for $p_e\!=\!0.8$, $P_{{av}}\!=\!1$ dB, $P_F\!=\!0.5$. As expected, $P_D$ increases as ${\cal K}$ increases and it saturates after certain ${\cal K}$, since $P_D$ is not limited by the battery size anymore and instead is limited by the sensors' statistics.
%In this Fig, by increasing ${\cal K}$, first $P_D$ is increasing function and then it saturated.
%
Comparing schemes I and II and their corresponding special cases in Figs. \ref{fig3-b} and \ref{fig3-c}, we make similar observations to those in Fig. \ref{fig3-a}.  

%%%
%\begin{figure}
    %\centering
    %\subfigure[]
    %{
        %\includegraphics[scale=.38]{PD_alpha_ave}\label{fig3-a}
    %}
    %\subfigure[]
    %{
        %\includegraphics[scale=.38]{PD_Pav_ave_db.eps}\label{fig3-b}
    %}
     %\subfigure[]
    %{
        %\includegraphics[scale=.38]{PD__k_ave.eps}\label{fig3-c}
    %}
    %\caption
    %{
        %(a) $P_D$ vs. $P_F$
        %(b) $P_D$ vs. $P_F$
        %(c) $P_D$ vs. ${\cal K}$
    %}
    %\label{result}
%\end{figure}
%%%%%%%%%%%%%%%%%%%%%%%%%%%%%%%%%%%%%%%%%%%%%%%%%%%%%%%%%%%%%%%%%%%%%%%%%%%%%%%%%%%%%%%%%%%
%\vspace{-.2cm}
%\section{Conclusion}
In summary, we studied a distributed detection problem in a wireless network with $K$ heterogeneous energy harvesting sensors and investigated the optimal local decision thresholds for given transmission and battery state models.
%
%As the optimality criterion, we considered the detection probability of the optimal fusion rule and two approximate expressions for the KL distance. %We derived the optimal fusion rule Considering the Kullback-Leibler divergence distance as a performance metric in scheme I, and $P_D$ in scheme II and applying energy and channel quality constrain, sensors decision rules are obtained. 
Our numerical results indicate that the thresholds obtained from maximizing the KL distance are near-optimal. Finding these thresholds is computationally very efficient, as it requires only $K$ one-dimensional searches, as opposed to a $K$-dimensional search required to find the thresholds that maximize the detection probability.
%
%Comparing the performance of scheme I with scheme II, illustrate that, both schemes have nearly the same. But optimization in scheme I has less complexity. Besides, It is shown that, considering different decision rule for each sensor has better performance. 
%Also, we formulated the optimal threshold of channel quality in decentralized energy harvesting sensor networks.
%%%%%%%%%%%%%%%%%%%%%%%%%%%%%%%%%%%%%%%%%%%%%%%%%%%%%%%%%%%%%%%%%%%%%%%%%%%%%%%%%%%%%%%%%%%%
\vspace{-.3cm}
\section*{Acknowledgment}
\vspace{-.1cm}
This research is supported by NSF under grant 1341966.
%%%%%%%%%%%%%%%%%%%%%%%%%%%%%%%%%%%%%%%%%%%%%%%%%%%%%%%%%%%%%%%%%%%%%%%%%%%%%%%%%%%%%%%%%%%
%\bibliographystyle{IEEEtran}
\bibliography{MyRef}

\end{document}